\documentclass[12pt]{article}
\usepackage{scicite}
\usepackage{times}

\usepackage{graphicx}
\usepackage{epsfig}
\newenvironment{sciabstract}{%
\begin{quote} \bf}
{\end{quote}}
\newcounter{lastnote}

\title{Encoding Complexity within Supramolecular Analogues of Frustrated Magnets}

\author
{Andrew B. Cairns,$^{1}$ Matthew J. Cliffe,$^{1}$ Joseph A. M. Paddison,$^{1,2}$\\
Dominik Daisenberger,$^3$ Matthew G. Tucker,$^{2,3}$ Fran{\c c}ois-Xavier Coudert,$^{4\ast}$\\
and Andrew L. Goodwin $^{1\ast}$ \\
\\
\normalsize{$^{1}$Department of Chemistry, University of Oxford, Inorganic Chemistry Laboratory,}\\
\normalsize{South Parks Road, Oxford OX1 3QR, U.K.}\\
\normalsize{$^{2}$ISIS Facility, Rutherford Appleton Laboratory, Harwell Oxford, Didcot,}\\
\normalsize{Oxfordshire OX11 0QX, U.K.}\\
\normalsize{$^{3}$Diamond Light Source, Chilton, Oxfordshire, OX11 0DE, U.K.}\\
\normalsize{$^{4}$institut de Recherce de Chimie Paris, CNRS -- Chimie ParisTech,}\\
\normalsize{11 rue Pierre et Marie Curie, 75005 Paris, France}\\
\\
\normalsize{$\ast$To whom correspondence should be addressed;}\\
\normalsize{E-mail: fx.coudert@chimie-paristech.fr, andrew.goodwin@chem.ox.ac.uk.}
}
\date{}

\begin{document} 

\baselineskip24pt

\maketitle 

\begin{sciabstract}
At the heart of systems chemistry lies the idea that supramolecular interactions can give rise to complex and unexpected collective states that emerge on a fundamentally different lengthscale to that of the interactions themselves \cite{r1,r2}. While in certain cases---\emph{e.g.}\ the self-assembly of virus-like polyhedral cages from coordination building blocks \cite{r3}---it is possible to control emergence in a systematic manner, the development of general approaches remains a fundamental challenge in the field \cite{r4}. In the conceptually-related domain of frustrated magnetism---where collective states give rise to exotic physics of relevance to data storage and spintronics---the task of predicting emergent behaviour is simplified through control over the geometry and form of the magnetic interactions from which complexity arises \cite{r5}. Seeking to combine approaches from these two fields, we study here the solid phases of inorganic polymer chains assembled from non-magnetic gold(I)/silver(I) cations and cyanide anions. We show the periodic inter-chain potential encodes a supramolecular interaction that can be tuned to mimic different magnetic interactions between XY spins (``spin rotors''). Because the chains pack on a triangular lattice, the crystal structures of gold(I)/silver(I) cyanides can be interpreted in terms of the phase behaviour of triangular XY magnets. Complex magnetic states predicted for this family---including hidden quadrupolar order and emergent spin-vortex quasiparticles \cite{r6}---are realised for the first time in the structural chemistry of these cyanide polymers. In this way we demonstrate both how simple inorganic materials might behave as structural analogues of otherwise-unrealisable ``toy'' spin models, and also how a theoretical understanding of those models might be used to predict and control emergent phenomena in chemical systems.
\end{sciabstract}

The existence of conceptual mappings between complex magnetic and structural states of matter has long been appreciated \cite{r7}. Arguably the most famous example is the low-temperature magnetic state of ``spin-ice'' Ho$_2$Ti$_2$O$_7$ \cite{r8}, in which tetrahedral spin clusters adopt a ``two- in-two-out'' arrangement geometrically related to the orientation of water molecules in cubic ice \cite{r9}. This mapping results in the same characteristically-large configurational entropy for both systems \cite{r10}, and even relates their excitations: the magnetic ``monopoles'' of spin ices and the Bjerrum defects of cubic ice are different manifestations of the same geometric defect \cite{r11}. Ice-like states are just one example of complex matter, and much of the field of frustrated magnetism is concerned with the extraordinary diversity of phases accessible through different specific combinations of lattice geometries, pairwise interactions, and effective degrees of freedom \cite{r12}. As in systems chemistry, the interest is very often in collective or emergent states where a system develops order on new length scales or evolves new degrees of freedom distinct to those of the individual components \cite{r13}. This interest is not only academic: topologically-protected emergent spin vortices in the skyrmion phases of chiral ferromagnets offer a possible mechanism of ultra-dense data storage \cite{r14}, and monopole quasiparticles in spin ices may yet be exploited in the emerging field of ``magnetronics'' \cite{r15}.

A natural question is whether chemical analogues might explore the same general phase space as frustrated magnets, perhaps allowing the realisation of systems that are difficult to access when considering magnetic interactions alone. For example, there are few magnetic systems that are strictly two-dimensional in nature, because some degree of interaction in a perpendicular direction usually persists \cite{r16}. Chemical analogues offer a neat solution to this problem: one-dimensional polymers can behave as single collective objects, such that the interactions between them are strictly two-dimensional. In this way, aggregates of polar polymer chains have provided some of the first physical realisations of the canonical triangular and square Ising lattices \cite{r17,r18,r19} and have been used to generate complex collective states based on geometric frustration \cite{r20}. Other aspects of mapping between magnetic and chemical states may prove more difficult to achieve. By their very nature, chemical systems lend themselves most naturally to discrete degrees of freedom---\emph{e.g.}, the switching of charge states, atom positions, or orbital orientations \cite{r7,r21,r22,r23}---and so it is not yet clear that many of the complex phases that require continuous degrees of freedom are actually accessible in a chemical context \cite{r24}.

We proceed to establish a mapping between the structural chemistry of gold(I)/silver(I) cyanides---compounds usually studied for their unusual thermal expansion behaviour \cite{r25,r26,r27}---and triangular magnets with continuous (XY) degrees of freedom, but first introduce the crystal structures of AuCN and AgCN themselves. In both cases, the transition metal cations are coordinated in a linear fashion by two cyanide ions, with each cyanide ion bridging two cations in turn \cite{r28}. Contrary to initial reports \cite{r29}, it is now established that there is no long-range order in the orientation of the cyanide ions for either system, so that both structures consist of essentially the same linear --M--(CN)--M-- chains of atoms \cite{r25,r26}. In each case these chains are arranged on a triangular lattice, but what differs between the two structures is the relative height of neighbouring chains: in the AuCN structure the chains are aligned such that Au$^+$ cations are in closest contact with other Au$^+$ cations, whereas in the AgCN structure there is a successive displacement by $\frac{1}{3}$ of the chain repeat length that brings Ag$^+$ cations closer to CN$^-$ ions than other Ag$^+$ cations [Fig.~1(a,b)]. It has been suggested that attractive metallophilic interactions (which are stronger for Au$^+$ than Ag$^+$ \cite{r30,r31,r32}) are sufficient to overcome electrostatics in AuCN but not in AgCN, so providing a qualitative understanding of the different structure types adopted \cite{r27,r33}. By associating the relative height $z_j$ of each M--(CN)--M chain with a phase angle $\theta_j=2\pi z_j$, we can map these structures onto triangular arrangements of XY spins $\mathbf S_j = S (\cos\theta_j,\sin\theta_j)$. In both cases each chain or spin has a single degree of freedom which is periodic in $z_j$ or $\theta_j$, respectively. The spin arrangements that correspond to the AuCN and AgCN structures are shown in Fig.~1(c,d); these are well known within the magnetism community as the ground states of the (nearest-neighbour) XY triangular ferromagnet and anti-ferromagnet, respectively \cite{r34}. A simple mapping between collective states in spin and cyanide structures is evident in the latter: the three-spin chiral degrees of freedom that characterise ``spiral order'' in the XY triangular antiferromagnet \cite{r35} correspond in the AgCN structure to argentophilic helices of a given handedness [Fig.~1(d)].

Our study begins by asking whether it is purely coincidental that these chemically-simple transition metal cyanides adopt structures that map onto the canonical XY triangular magnets? We have already highlighted the clear relationship that exists between the lattice geometry and effective degrees of freedom in both cases, but to answer this question properly we need to understand how and why the pairwise interactions are also related. To this end, we have used quantum mechanical calculations to determine the form of the supramolecular interaction potential operating on pairs of neighbouring MCN chains as a function of their relative phase $\Delta\theta=2\pi\Delta z$ [Fig.~2(a)]. We might expect two dominant contributions to this potential, with the particular balance observed depending on the chemistry of the relevant system: metallophilic interactions will favour coalignment of chains in order to minimise M$\ldots$M separations, whereas electrostatics will favour staggering so as to bring the M$^+$ cations of one chain into registry with the CN$^-$ anions of the neighbouring chain. Our computational results are shown in Fig.~2(b), from which it is clear that metallophilicity dominates in the case of AuCN (energy minimum at $\Delta\theta=0$) but not in the case of AgCN (energy minimum at $\Delta\theta=\pm\pi$). Crucially, in both cases, the potentials are well approximated by their first-order Fourier component
\begin{equation}
E=J\cos(\Delta\theta),
\end{equation}
with $J = -7.03(18)$\,kJ\,mol$^{-1}$ and $+3.71(5)$\,kJ\,mol$^{-1}$ for AuCN and AgCN, respectively [Fig.~2(b)].

The similarity to the spin Hamiltonian for the XY triangular (anti)ferromagnet
\begin{equation}
\mathcal H=J_{\rm{XY}}\sum_{i,j}\mathbf S_i\cdot\mathbf S_j=J_{\rm{XY}}S^2\sum_{i,j}\cos(\Delta\theta_{ij})
\end{equation}
is striking (here, $i, j$ index neighbouring spins, and $\Delta\theta_{ij}$ describes the angle between spins $i$ and $j$) and suggests that we can understand the structural chemistry of both cyanides in terms of XY spins that interact via an effective nearest-neighbour exchange constant $J_{\rm{XY}} = J/S^2$. So, for AuCN it is because $J < 0$ that its crystal structure maps onto the ground state of the XY triangular ferromagnet ($J_{\rm{XY}} < 0$); conversely, the positive value of $J$ for AgCN explains why its structure is related to the ground state of the XY triangular antiferromagnet. We note that in both cases the effective exchange energies are larger than $k_{\rm B}T$ under ambient conditions and orders of magnitude larger than in conventional magnetic systems.

From a supramolecular chemistry perspective, what we are saying is that the self-interaction potentials of AuCN and AgCN chains encode for different collective structures, much as the geometries of coordination chemistry building blocks might encode for differently-shaped polyhedral assemblies \cite{r3}. For AuCN, energy is minimised by bringing chains into alignment, resulting in the simple layered structure illustrated in Fig.~1(a) and the X-ray diffraction pattern shown in Fig.~2(c). For AgCN, pairwise interactions are minimised by staggering neighbouring chains by one half of the chain repeat length. This local arrangement cannot be propagated on the triangular lattice (since one pair of any triplet of neighbouring chains would be forced to adopt the same height); instead the lowest-energy compromise \cite{r34} is for successive heights to differ by $\Delta z=\frac{1}{3}$, giving the structure represented in Fig.~1(b) and the X-ray diffraction pattern shown in Fig.~2(c). In both cases the higher-order Fourier components to the real interaction potential [Fig.~2(a)] will influence subtle features of the structures; for example, there is likely to be some buckling of the cation layers in AuCN that arises because the energy minima occur for small but nonzero values of $\Delta\theta$. Nevertheless, the important structural features are clearly captured by the simple model presented here. The mapping onto XY spin systems serves two purposes. On the one hand, it establishes physical realisations of these theoretical spin models which in principle enable aspects of the models to be tested and studied experimentally. (Note that three-dimensional realisations of triangular magnets generally differ in their behaviour relative to the theoretical two-dimensional case \cite{r36}.) On the other hand, our understanding of the theory of these magnets allows us to interpret and predict structural features (such as the ground state) of the experimental system themselves.

Having established a simple mapping for AuCN and AgCN onto the canonical XY triangular magnets, we sought to understand the structural chemistry of the more complex heterometallic phase Ag$_{1/2}$Au$_{1/2}$(CN). By way of background, this system is known to adopt a similar triangular rod packing to the homometallic cyanides, with (N-bound) Ag and (C-bound) Au atoms
now alternating strictly along any given chain [Fig.~3(a)] \cite{r27}. Despite this local order, the arrangement of chains relative to one another remains uncertain and the true crystal structure is not known \cite{r27}. What is clear---both from the data of Ref. \citenum{r27} and our own measurements (see SI)---is that the X-ray diffraction pattern contains regions of structured diffuse scattering that are the signature of the correlated disorder associated with complex states [Fig.~3(b)] \cite{r37}. Our approach to studying this system is as follows: (i) establish, using quantum chemistry methods, the form of the self-interaction potential for Au--CN--Ag--NC--Au chains, (ii) determine a mapping onto a suitable XY triangular magnet, and (iii) seek to interpret the experimental diffraction pattern of Ag$_{1/2}$Au$_{1/2}$(CN) in terms of a structural model determined by analogy to the corresponding magnetic system. In developing this mapping we note the doubled repeat length of Au--CN--Ag--NC--Au chains relative to Au--(CN)--Au and Ag--(CN)--Ag means that the phase shift $\Delta\theta$ adopts a subtly different meaning in this more complicated case. In particular, both $\Delta\theta=0$ and $\Delta\theta=\pi$ now correspond to situations with strong metallophilic interactions, differing only in whether like ($\Delta\theta=0$) or unlike ($\Delta\theta=\pi$) atoms are in close contact [Fig.~3(a)].

Quantum mechanical calculations of the chain self-interaction potential give the trend shown in Fig.~3(c). This energy profile takes a more complex form than for the homometallic cyanides in part because of the Ag/Au alternation and in part because this alternation is coupled to CN orientational ordering. In simplifying the function $E(\Delta\theta)$ it is now appropriate to consider the first two Fourier components, so as to account for the difference in average metallophilic and electrostatic interactions (each with periodicity $\Delta\theta=\pi$) as well as the difference in homometallic and heterometallic interactions (periodicity $\Delta\theta=2\pi$):
\begin{eqnarray}
E&=&J_1\cos(\Delta\theta)+\frac{J_2}{2}\cos(2\Delta\theta)\\
&\equiv&J_1\cos(\Delta\theta)+J_2\cos^2(\Delta\theta).\label{j2}
\end{eqnarray}
This approximation is poorest at the highest-energy region of the interaction potential, but accounts reasonably well for the overall shape near the energy minimum, the more critical feature [Fig.~3(c)]. Our data give $J_1 = +2.3(3)$\,kJ\,mol$^{-1}$ and $J_2 = -1.3(5)$\,kJ\,mol$^{-1}$. The $J_2$ term describes the difference in average metallophilic and electrostatic interactions, and its small value demonstrates the especially delicate balance between these two terms in this case---this is no surprise, of course, given the different structures of AgCN and AuCN themselves. Instead the dominant interaction in Eq.~(\ref{j2}) is described by the $J_1$ term; its positive value corresponds to a preference for heterometallic Ag$\ldots$Au contacts relative to a combination of homometallic Ag$\ldots$Ag and Au$\ldots$Au contacts.

The relevant magnetic system is now the so-called bilinear-biquadratic (BLBQ) XY model, characterised by the spin Hamiltonian
\begin{eqnarray}
\mathcal H&=&J_1\sum_{i,j}\mathbf S_i\cdot\mathbf S_j+J_2\sum_{i,j}|\mathbf S_i\cdot\mathbf S_j|^2\\
&=&J_1\sum_{i,j}\cos(\Delta\theta_{ij})+J_2\sum_{i,j}\cos^2(\Delta\theta_{ij}),\label{e6}
\end{eqnarray}
where we have subsumed the dependency on spin magnitude $S$ within the $J_n$. In practice, the BLBQ model is most frequently used to describe $S = 1$ systems \cite{r38,r39}, with $J_1$ and $J_2$ terms quantifying the strength and nature of dipolar and quadrupolar interactions, respectively. For XY spins, the thermodynamic phase diagram is known from theory \cite{r6}; the particular combination of $J_1,J_2$ values determined above corresponds to a complex ``nematic'' state in which dipolar disorder is coupled with (hidden) ferroquadrupolar order [Fig.~3(d)]. To the best of our knowledge, this state has never been realised in a physical system \cite{r16,r38,r40}. The implication for the structure of Ag$_{1/2}$Au$_{1/2}$(CN) is that chains should align to bring metal cations in registry with one another, but with a preference for unlike Ag$\ldots$Au neighbours. This arrangement is geometrically frustrated (because any triplet of neighbouring Ag/Au atoms must contain one pair of like atoms), and would represent a disordered state that is essentially a superposition of the family of acceptable structure solutions proposed in Ref. \citenum{r27} on the basis of their fits to neutron pair distribution functions. The X-ray diffraction pattern calculated for a structural model of Ag$_{1/2}$Au$_{1/2}$(CN) based on this strict quadrupolar order is shown in Fig.~3(b). While the general intensity distribution and reflection conditions of the diffraction pattern are reproduced, the degree of diffuse scattering and selective peak broadening is underestimated. (Note that equivalent calculations based on the other phases described by the $J_1, J_2$ phase diagram are clearly inconsistent with experiment, see SI). We proceed to show that the discrepancy in peak widths can be accounted for by considering finite-temperature effects in the XY model.

It has long been known that triangular XY magnets are unstable at $T > 0$ with respect to the spontaneous formation of spin vortices, which act to broaden Bragg reflections in the corresponding magnetic scattering patterns \cite{r41}. These vortices are topologically-protected emergent quasi-particles with their own peculiar physics---\emph{e.g.}\ the so-called ``Kosterlitz--Thouless'' transition that allows vortex unpinning at elevated temperatures \cite{r42}---and are conceptually related to the magnetic skyrmion states of interest in data storage applications \cite{r43}. Monte Carlo simulations of two-dimensional spin states are known to be extremely sensitive to finite size effects \cite{r44,r45}, but we find that a simple quench Monte Carlo protocol driven by Eq. (\ref{e6}) yields conveniently-sized configurations that capture the two essential ingredients of interest to our current study: namely, quadrupolar order and spin vortices [Fig.~3(e)] (see SI). We interpret these configurations as non-equilibrium approximations to the computationally-inaccessible finite-temperature (equilibrium) spin structure. The corresponding structural model for Ag$_{1/2}$Au$_{1/2}$(CN), represented in Fig.~3(f), finally yields an X-ray diffraction pattern with the correct variation in both intensity and peak widths [Fig.~3(b)]. Consequently we can be confident that this model---which is the first ever to reproduce the experimental diffraction pattern \cite{r27}---captures the key structural features of Ag$_{1/2}$Au$_{1/2}$(CN). The subtleties in energy profile corresponding to higher-order Fourier terms [Fig.~3(c)] will likely result in small additional modulations of the metallophilic layers, but clearly these are not required to account for the diffraction behaviour. Our mapping between spin states and chain configurations has translated spin vortices into screw dislocations, which have become the topologically-protected emergent objects responsible for the experimentally-observed selective peak broadening [Fig.~3(f)]. Remarkably, our analysis implies that these collective multi-chain entities are actually encoded for by the self-interaction potential of an individual --Au--CN--Ag--NC--Au-- chain. Looking further ahead, this link between individual chemical component and large-scale topological defect suggests a broader supramolecular strategy for engineering complex materials; \emph{e.g.}\ the controlled incorporation of screw dislocations as a means of tuning the electronic band structure of topological insulators \cite{r46}.

While we have focussed on interpreting the structural consequences of this mapping between supramolecular and magnetic interactions, there is of course scope for drawing similar parallels between excitations and/or the response to external stimuli ($T,p,H,\ldots$) of both systems \cite{r47}. Are the screw dislocations we observe pinned or mobile, for example? Can temperature or pressure be used to vary the relative strengths of the different exchange parameters? There is cause here for optimism: a preliminary investigation of the high-pressure behaviour of AgCN (see SI) suggests slow transformation to an AuCN-structured polymorph that might be rationalised in terms of an inversion of $J_{\rm{eff}}$ as argentophilic interaction energies increase more rapidly with decreasing inter-chain separation than the electrostatic contribution ($\propto r^{-6}$ \emph{vs} $r^{-1}$) \cite{r32}. Given the even more delicate balance of competing energy terms observed in Ag$_{1/2}$Au$_{1/2}$(CN) and the accessibility of extensive chemical substitution \cite{r27}, variable-temperature/pressure studies of the heterometallic cyanides likely offer an especially fruitful avenue of future research. In general, one might expect that kinetics will play a greater role for supramolecular analogues of frustrated magnets than in the magnetic systems themselves, given that the effective exchange energies involved are typically larger by one or two orders of magnitude. On the one hand, this has the likely advantage of stabilising the most interesting phases to ambient conditions---as is necessary, for instance, if topological states are actually to be employed in room-temperature data storage technology. But, on the other hand, the key disadvantage will be the difficulty faced in interpreting phase behaviour in terms of equilibrium models, which are arguably the mainstay of the field of frustrated magnetism. The presence of multiple local minima in the self-interaction potential (\emph{e.g.}\ as seen in Fig.~3(c)) may add further complications. One way or the other, what we have demonstrated here is how the interaction potential encoded within a material fragment might be tuned chemically to give rise to a variety of structural phases---both simple and complex---the nature of which can be reliably predicted by exploiting a mapping onto geometrically-related problems in the seemingly-unrelated field of frustrated magnetism.

\section*{Methods}

\subsection*{Quantum mechanical calculations}

Quantum chemical calculations were performed in order to measure the variations in enthalpy in pairs of metal cyanide chains as a function of phase shift $\Delta\theta$. For each system, two periodic chains were considered, with their equilibrium geometry and a interchain distance fixed to the crystallographic values. The chains were then shifted by increments of $2^\circ$, and for each value of the shift a single-point energy calculation was performed, resulting in $\Delta E(\Delta\theta)$ profiles. In the case of the homometallic AgCN and AuCN chains, where there is no order in the orientation of the cyanide ions, we used for the chains a repeating unit equal to four times the basic --M--(CN)-- motif, in order to account for all four possible orientations of the two cyanide anions

All calculations were performed in the density functional theory approach with localised basis sets, using the CRYSTAL14 software \cite{r48}. We used all-electron Triple-Zeta Valence with Polarisation basis sets for C and N atoms \cite{r49}, an energy-consistent relativistic 19-valence-electron pseudopotential and corresponding valence basis set for Ag \cite{r50}, and an effective core potential for Au \cite{r51}. The shrinking factor of the reciprocal space net is set to 8, corresponding to 65 reciprocal space points at which the Hamiltonian matrix was diagonalised. The total energy obtained with this mesh is fully converged.

We used the PBEsol0 global hybrid exchange--correlation functional \cite{r52,r53}: of the exchange--correlation functionals benchmarked on the bulk structures of AgCN and AuCN, it gave the best agreement with experimental lattice parameters. Similarly, we investigated the effect of empirical dispersion corrections (in the Grimme scheme \cite{r54}) and showed that they did not change the conclusions reached (\emph{i.e.}\ the relative stability of the different structures studied). However, because the Ag--Ag and Au--Au interatomic distances in this study are close to the cutoff of the empirical correction, their inclusion might introduce non-physical behaviour in the energy profiles. As a consequence, all results presented here were obtained without any empirical correction for dispersion.

\subsection*{X-ray powder diffraction calculations}

X-ray powder diffraction patterns were calculated using CrystalDiffract software. Peak shapes were modelled using a Gaussian function of empircally-determined width $0.4^\circ$. Corrections for preferred orientation were implemented, with the relevant parameters determined by inspection of the experimental data in Ref.~\citenum{r27}: a plate-like geometry was used for AuCN and Ag$_{1/2}$Au$_{1/2}$(CN) patterns (alignment parameter 0.47) and a needle-like geometry was used for AgCN (alignment parameter 0.29).

\cleardoublepage
\baselineskip24pt

%authors list, title of paper, name of journal in abbreviation, volume number, initial-final page number (or article number), (year).

\section*{Acknowledgements}
A.B.C., M.J.C., J.A.M.P. and A.L.G. gratefully acknowledge financial support from the E.P.S.R.C. (EP/G004528/2), S.T.F.C. and the E.R.C. (Grant Ref: 279705). High-pressure synchrotron X- ray powder diffraction measurements were carried out at the I15 beamline, Diamond Light Source, U.K. Quantum chemistry calculations made use of HPC resources from GENCI (Grant x2015087069).

\clearpage

\noindent {\bf Fig.\ 1.} \\
{\bf Crystal structures of AuCN and AgCN and their relationship to the ground states of triangular XY (anti)ferromagnets.} The crystal structures of both (a) AuCN and (b) AgCN consist of one-dimensional metal--cyanide--metal chains arranged on a triangular lattice. The key difference between the two structures concerns the relative heights of neighbouring chains. In AuCN, the metal atoms (large spheres) form close-packed planes; this arrangement has hexagonal $P6/mmm$ symmetry. In AgCN, neighbouring chains are displaced by $\frac{1}{3}$ of the chain repeat so as to bring metal cations in closer contact with the cyanide anions (smaller spheres). This structure has a larger unit cell with trigonal $R\bar3m$ symmetry. The relative height of each chain can be represented by a phase angle, shown here projected onto the underlying triangular lattice. The magnetic ground states of the triangular XY (c) ferromagnet and (d) antiferromagnet correspond directly to this phase interpretation of the structures of AuCN and AgCN. In each case the relevant unit cell is outlined in bold. The chiral degrees of freedom in the ``spin spiral'' antiferromagnetic state---given by the sense of spin rotation as the vertices of a triangular plaquette (shaded) are traversed in a common direction---translate to the handedness of argentophilic helices in the AgCN structure (inset to panel (d)).
\clearpage

\noindent {\bf Fig.\ 2.} \\
{\bf Interactions in AuCN and AgCN.} (a) The relative displacement between neighbouring chains, $\Delta z$, can be mapped uniquely onto a phase shift $\Delta\theta$ between corresponding XY spins. (b) The variation in enthalpy for a pair of (top) AuCN and (bottom) AgCN chains, separated by their respective equilibrium distances, calculated as a function of phase shift $\Delta\theta$. Data are shown as filled circles, and the first-order Fourier component is shown as a red line. (c) Comparison of experimental \cite{r27} and calculated X-ray diffraction patterns (Cu K$\alpha$ radiation) for (top) AuCN and (bottom) AgCN. The structural models used for both calculations were derived from the ground state of the triangular XY magnets to which the relevant energy profile in (b) corresponds.

\clearpage

\noindent {\bf Fig.\ 3.} \\
{\bf Interactions and structural complexity in Au$_{1/2}$Ag$_{1/2}$(CN).} (a) The metal--cyanide chains in Au$_{1/2}$Ag$_{1/2}$(CN) consist of strictly-alternating Au--CN--Ag--NC--Au linkages (gold atoms in yellow; silver atoms in grey). Consequently the chain repeat length is approximately twice that in AgCN or AuCN. In mapping the relative heights of neighbouring chains $\Delta z_{ij}$ onto phase shifts $\Delta\theta_{ij}$ , the meaning of the $\Delta\theta_{ij}$ is now subtly different to that in Fig.~2; some representative cases are illustrated here. (b) A comparison of the experimental X-ray diffraction pattern (Cu K$\alpha$ radiation) of Au$_{1/2}$Ag$_{1/2}$(CN) \cite{r27} with that calculated for the ground-state (blue) and finite-temperature (red) models described in the text. (c) The variation in enthalpy for a pair of Au$_{1/2}$Ag$_{1/2}$(CN) chains, separated by their respective equilibrium distances, calculated as a function of phase shift $\Delta\theta$. Data are shown as filled circles, and the second-order Fourier component is shown as a red line. (d) The equilibrium phase diagram for the BLBQ triangular XY model, adapted from Ref.~\citenum{r6}. The parameters $J_1, J_2$ determined from the fit shown in panel (c) correspond to the point marked by a green star, which sits within the ferroquadrupolar (FQ) region of the phase diagram. (e) A representative Monte Carlo spin configuration generated as described in the text; here different colours correspond to different spin orientations within the plane (inset). Ferroquadrupolar order is evident in the existence of local ``stripe'' patterns. (f) A section of the structural model for Au$_{1/2}$Ag$_{1/2}$(CN) generated directly from the spin configuration shown in panel (e). Note the absence of Ag/Au order within the metal-containing layers. (g) A section of the same model, with C and N atoms omitted and three screw dislocations highlighted in red.

\clearpage

\begin{center}
\includegraphics[width=8.3cm]{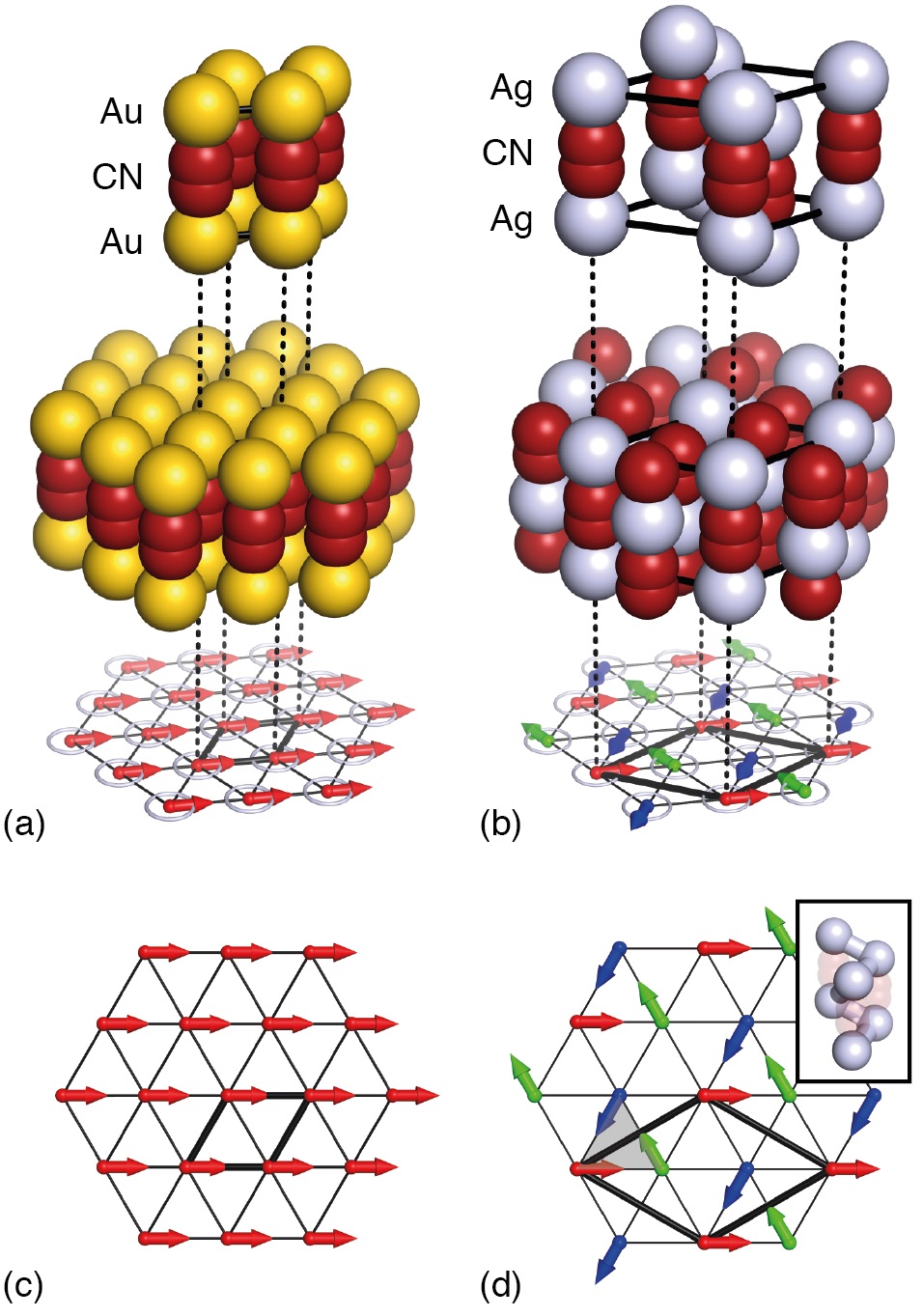}\\
FIGURE 1\\
\end{center}
\clearpage

\begin{center}
\includegraphics[width=8.3cm]{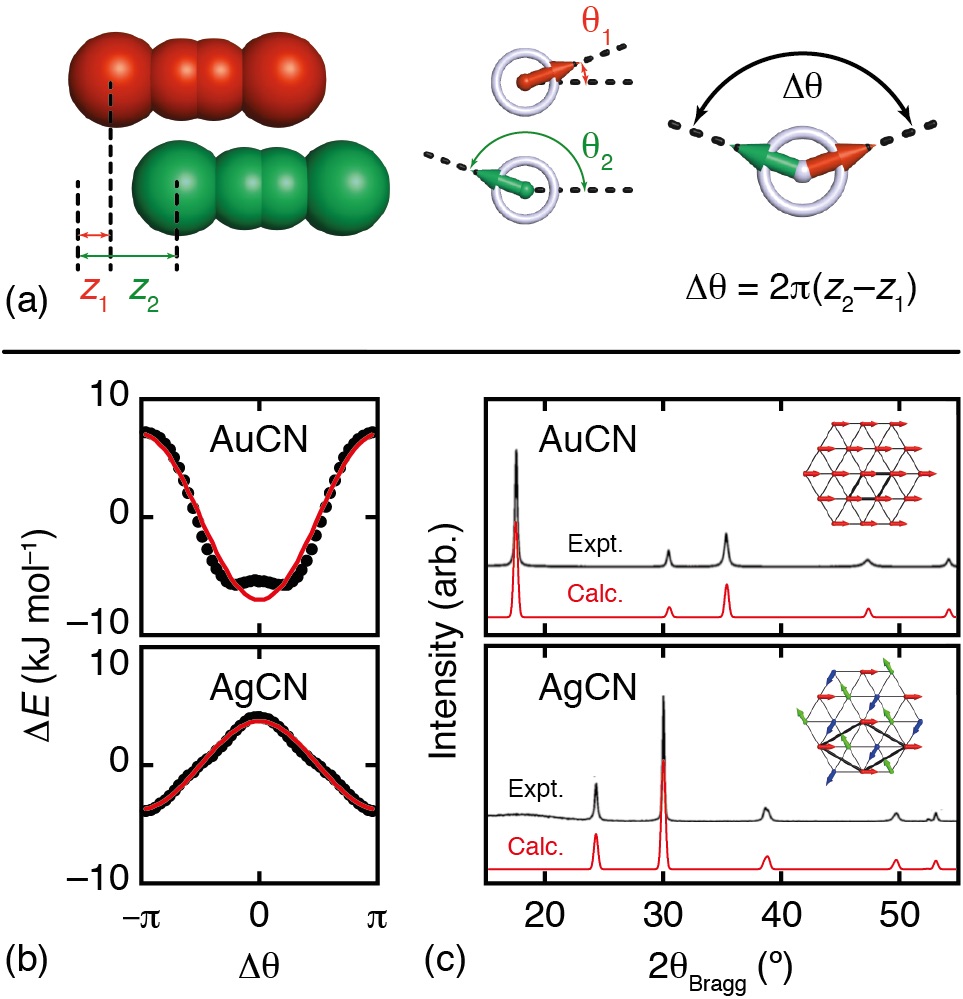}\\
FIGURE 2\\
\end{center}
\clearpage

\begin{center}
\includegraphics[width=8.3cm]{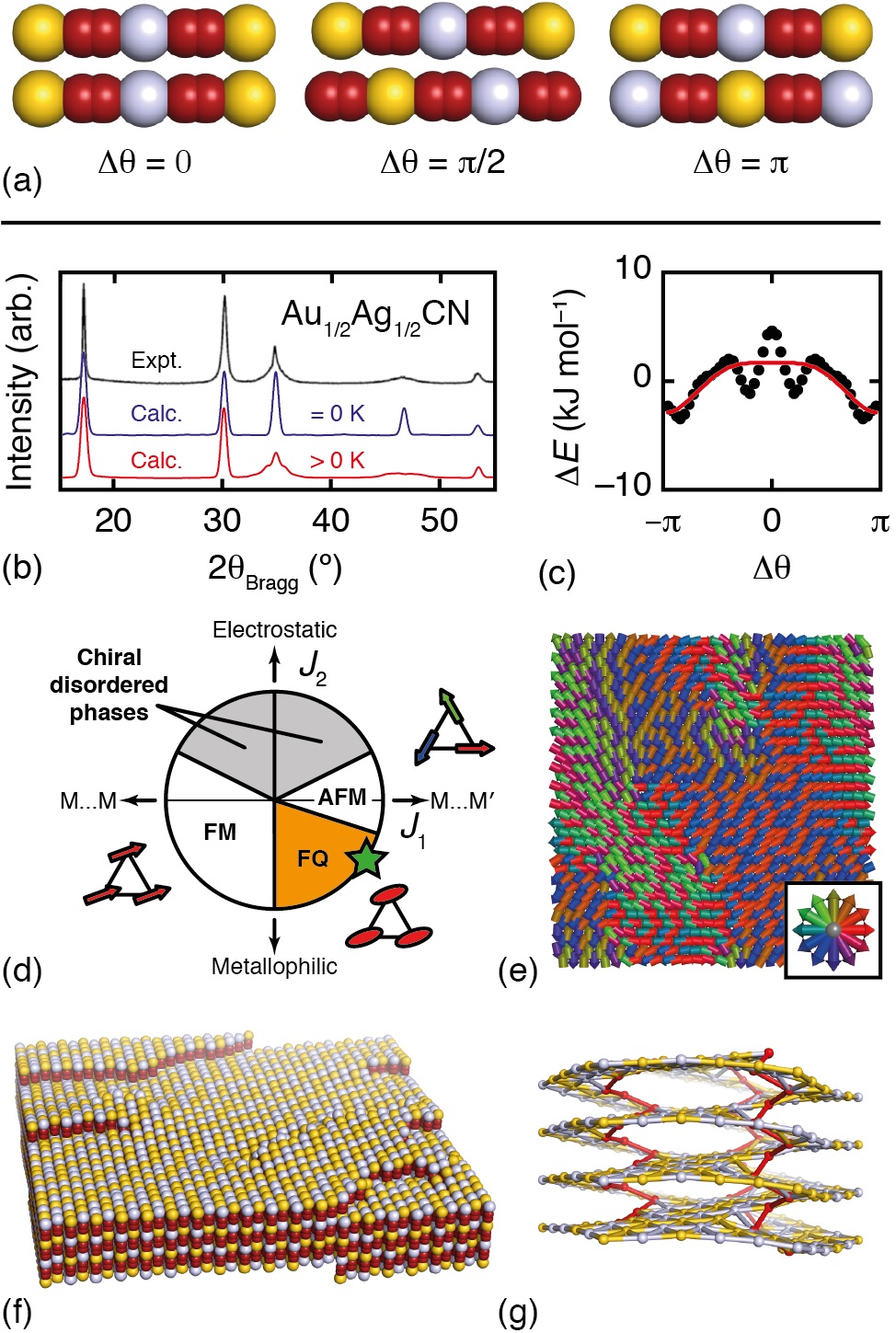}\\
FIGURE 3\\
\end{center}

\end{document}